\documentclass[preprint,showkeys,secnumarabic,amsfonts,showpacs,amsmath,amssymb]{revtex4}
\usepackage[dvips]{color}
\usepackage{epsfig}
\usepackage{amsmath}
\usepackage{graphicx}

\begin{document}
\title{Dynamics of the universe as a "test 3-brane" in a 5D bulk  }

\author{Hossein Farajollahi}
\email{hosseinf@guilan.ac.ir} \affiliation{Department of Physics,
University of Guilan, Rasht, Iran}
\author{Arvin Ravanpak}
\email{aravanpak@guilan.ac.ir} \affiliation{Department of Physics,
University of Guilan, Rasht, Iran}

\date{\today}

\begin{abstract}
In this paper we study the dynamics of a 5D bulk space-time with our universe as a "test
3-brane" located in the bulk, an idealized model of a topological object whose physical properties
is negligible in comparison with that of the bulk. Our universe experiences acceleration and its equation of state parameter crosses the phantom divide line due to the geometry of the bulk space-time.\\

\end{abstract}



\maketitle

\section{Introduction}
Cosmological observations, such as Super-Nova Ia (SNIa), Wilkinson Microwave
Anisotropy Probe (WMAP), Sloan Digital Sky Survey (SDSS), Chandra X-ray Observatory
 etc. reveal some cross-checked information of our universe that support an acceleration of the universe expansion\cite{obs00}--\cite{obs4}. Dark energy (DE)
component is a candidate to drive the acceleration in addition to represent a dynamical EoS model for the universe \cite{dark1}--\cite{eos4}.

Alternatively, the higher dimensional theories may explain cosmic acceleration and phantom crossing \cite{mtheory7}\cite{mtheory}. The existence of extra dimensions is required in various theories beyond the standard model of particle physics, especially
in theories unifying gravity with the other fundamental forces, such as superstring and M theories \cite{mtheory3}--\cite{mtheory6}. In brane cosmological models our 4D universe would be a surface or a 3-brane embedded into a higher dimensional bulk space-time on
which gravity could propagate and energy exchanges between
the brane and the bulk  \cite{acc1}--{\cite{mod1}. In these models,  an infrared modification of general
relativity on large scales, by weakening the gravitational interaction on those scales, may allow the recent acceleration
expansion of the universe. This idea is motivated by the fact that our observations of
gravity is from sub-millimiter scales up to solar system scales while the Hubble radius is many orders of magnitude larger \cite{brane1}--\cite{brane4}. Among these, the RS brane-world models inspired by D-brane geometry in string theory and ,in particular, the RS II model is very popular since it has a new modification of the gravitational potential in the very early stages of the universe evolution \cite{brane2}--\cite{bra2}}. The brane approach provides a new way of understanding the hierarchy between the 4D Planck scale and the electro-weak scale that
leads to constructing models that exhibit crossing of the phantom divide line \cite{crossing}--\cite{bounce3}. One common feature in all these models is that by imposing the Israel junction condition which relates the jumps of the derivative of the metric across the brane to the stress-energy tensor inside the brane, the effective field equations in the brane are obtained \cite{israel2}. However, in this work, we assume that the branes are "test 3-branes" in the bulk similar to the idea of "test particles" in a gravitation field in 4D spacetime. The "test 3-branes" contain no matter of any kind with no tension and so the energy momentum tensor in the branes vanishes. Thus, there is no discontinuities of the metric coefficients ( or more precisely of their derivatives in the fifth direction normal to the brane). The price we pay
for this simplification is that in general we do not have field equations for the branes. Instead, we have enough information to predict the bevaiour of the test 3-brane affected by the dynamics of the bulk.

\section{The model}

We consider two 3-branes, described by 4D hyper-surfaces, embedded in a 5D bulk space-time, with the action given by
\begin{equation}\label{action}
S=\frac{1}{\kappa_5}\int_{bulk} d^5x\sqrt{-^{(5)}g}(^{(5)}R)+\frac{1}{\kappa_5}\sum_{i=1,2}\int_{brane} d^4x\sqrt{-g^{(i)}}K_i,
\end{equation}
where $\kappa_5$ is the 5D gravitational constant, $^{(5)}R$ is the 5D Ricci scalar and $K_i, \ \ i=1,2$ are the extrinsic curvatures of the branes in the higher dimensional bulk. In the above we assumed 1) a pure bulk with no cosmological constant and matter and 2) two "test 3-branes" (one of them is our universe) defined by $y=y_1$ and  $y=y_2$ with no tension and matter. In the DGP model, the brane tension and the bulk cosmological constant are both set to
zero, and the current acceleration of the universe expansion is explained as an effect of extra
dimension. In  here, we do not intend to derive the effective 4D Einstein equations
by projecting the 5D metric onto the brane. Instead, we would like to obtain the field equations in a general form in 5D bulk.

The most general 5D metric for homogeneous and isotropic universes with respect to the spatial coordinates $x^\mu$ is given by \cite{crs}--\cite{Cline}
\begin{equation}\label{met}
ds^2=-n(t,y)^2dt^2+a(t,y)^2\gamma_{\mu\nu}dx^\mu dx^\nu+2c(t,y)dtdy+b(t,y)^2dy^2,
\end{equation}
where $\gamma_{\mu\nu}$ is a maximally symmetric 3D metric. We consider a flat FRW metric for the 3D metric and without lose of generality, we adopt Gaussian normal coordinates, where $(c = 0)$.

 The field equations in 5D are
\begin{eqnarray}
  && 3[-\frac{n^2}{b^2}(\frac{a''}{a}+\frac{a'}{a}(\frac{a'}{a}-\frac{b'}{b}))+\frac{\dot{a}}{a}
(\frac{\dot{a}}{a}+\frac{\dot{b}}{b})]=0, \label{Gtt}\\
 && \frac{a^2}{b^2}[2\frac{a''}{a}+\frac{n''}{n}+\frac{a'}{a}(\frac{a'}{a}+2\frac{n'}{n})
-\frac{b'}{b}(\frac{n'}{n}+2\frac{a'}{a})] \nonumber \\& & -\frac{a^2}{n^2}[2\frac{\ddot{a}}{a}+
\frac{\ddot{b}}{b}+\frac{\dot{a}}{a}(\frac{\dot{a}}{a}-2\frac{\dot{n}}{n})+\frac{\dot{b}}{b}
(2\frac{\dot{a}}{a}-\frac{\dot{n}}{n})]=0, \label{Gii}\\
&& 3[\frac{a'}{a}(\frac{a'}{a}+\frac{n'}{n})-\frac{b^2}{n^2}(\frac{\ddot{a}}{a}+
\frac{\dot{a}}{a}(\frac{\dot{a}}{a}-\frac{\dot{n}}{n}))]=0, \label{Gyy}\\
 && 3[\frac{n'}{n}\frac{\dot{a}}{a}+\frac{a'}{a}\frac{\dot{b}}{b}-\frac{\dot{a}'}{a}]=0, \label{Gty}
\end{eqnarray}
where dot and prime respectively represent derivatives with respect to time $t$ and extra dimension $y$. In the above we consider
the two parallel branes at the endpoints of the y-coordinate. Let us now consider the Einstein equations (\ref{Gtt})--(\ref{Gty}) by considering the proper transverse distance between the
branes is time independent, a static fifth dimension, or, $\dot{b} = 0$ and for simplicity take $b = 1$. The motivation for considering a static fifth dimension is discussed by the authors in \cite{Arnowitt}--\cite{Leon}. In particular it has been argued that a static fifth dimension does not allow the inclusion
of brane matter and is a requirement of Randall-Sundrum-type solutions \cite{csaki}--\cite{Mohapatra}. The equations (\ref{Gtt})--(\ref{Gyy}) then become
\begin{eqnarray}
&& \frac{a''}{a}+\frac{a'^2}{a^2}-\frac{1}{n^2}\frac{\dot{a}^2}{a^2}=0, \label{A}\\
 && 2\frac{a''}{a}+\frac{n''}{n}+\frac{a'}{a}(\frac{a'}{a}+2\frac{n'}{n})-\frac{1}{n^2}[2\frac{\ddot{a}}{a}
+\frac{\dot{a}}{a}(\frac{\dot{a}}{a}-2\frac{\dot{n}}{n})]=0, \label{B} \\
&& \frac{a'}{a}(\frac{a'}{a}+\frac{n'}{n})-\frac{1}{n^2}[\frac{\ddot{a}}{a}
+\frac{\dot{a}}{a}(\frac{\dot{a}}{a}-\frac{\dot{n}}{n})]=0. \label{C}
\end{eqnarray}
In terms of the Hubble parameter by defining  $H_t\equiv\frac{\dot{a}}{a}$  and  $H_y\equiv\frac{a'}{a}$, the above equations can be written as
\begin{eqnarray}
  && H_t^2 = n^2[H'_y+2H_y^2], \label{a1} \\
  && 2\dot{H_t}+3{H_t}^2 = n^2[2H'_y+3H^2_y+\frac{n''}{n}+2H_y\frac{n'}{n}+2H_t\frac{\dot{n}}{n^3}],\label{b} \\
&&  H_y^2+H_y\frac{n'}{n}-\frac{\dot{H_t}}{n^2}-2\frac{H_t^2}{n^2}+H_t\frac{\dot{n}}{n^3}=0. \label{c}
\end{eqnarray}
Comparing the above equations with the standard Friedmann equations we obtain the generalized energy density and pressure, $\rho_{g}$ and  $p_{g}$ as
\begin{eqnarray}
\rho_{g} &=& 3n^2[H'_y+2H_y^2]. \label{rho1}\\
  p_{g} &=& -n^2[2H'_y+3H^2_y+\frac{n''}{n}+2H_y\frac{n'}{n}+2H_t\frac{\dot{n}}{n^3}].\label{p1}
\end{eqnarray}
Using equations (\ref{rho1}) and (\ref{p1}), the conservation equation can be written in the standard form as
\begin{equation}\label{eos}
\dot{\rho}_{g}+3H_t(1+\omega_{g})\rho_{g}=0,
\end{equation}
where the generalized EoS parameter, $\omega_{g}$, in the bulk for the model is given by
\begin{equation}\label{eos1}
\omega_{g}=-1+\frac{-H'_y-3H^2_y+\frac{n''}{n}+2H_y\frac{n'}{n}+2H_t\frac{\dot{n}}{n^3}}{3(H'_y+2H_y^2)}.
\end{equation}
Obviously, for the generalized EoS parameter to cross the phantom divide line, when $\omega_{g}\rightarrow -1$, we have: 1) $H_t\neq 0$, 2) $\dot{H_t}=\frac{-1}{2}\rho_{g}(1+\omega_{g}) \rightarrow 0$, 3) $\omega_g$ has different signs before and after crossing and $-1$ at the time of crossing, and 4) $2\ddot{H_t}=-\frac{d}{dt}(\rho_{g}+p_{g})\neq 0$.

By using equations (\ref{a1})-(\ref{p1}) we have

\begin{eqnarray}
H_t &=& \pm n\sqrt{H'_y+2H_y^2},\label{H_t}\\
\dot{H_t}&=&-\frac{n^2}{2}[H'_y+3H^2_y-\frac{n''}{n}-2H_y\frac{n'}{n}-2H_t\frac{\dot{n}}{n^3}],\label{dH_t}\\
 \ddot{H_t}&=& H_t[\frac{\ddot{n}}{n}-\frac{\dot{n^2}}{n^2}]+n'[H_y\dot{n}+n\dot{H_y}]+H_y[n\dot{n}'-3n^2\dot{H_y}-3n\dot{n}H_y] +\nonumber \\& & \dot{n}[\frac{n''}{2}+\frac{\dot{H_t}}{n}-nH_y']+\frac{n^2}{2}[\frac{\dot{n}''}{n}-\dot{H_y}'],\label{ddH_t}
\end{eqnarray}

where for an expanding universe we take the positive sign of $H_t$. One can easily find that from (\ref{H_t})--(\ref{ddH_t}) to satisfy the above the above conditions, we have 1) $n\neq 0$ and $H_y\neq\frac{1}{2y+C(t)}$, 2) $H'_y+3H^2_y=\frac{n''}{n}+2H_y\frac{n'}{n}+2H_t\frac{\dot{n}}{n^3}$ and 3) $\ddot{H_t}\neq 0$.
Note that in a normal treatment of gauge models in physics, we usually fix the gauge in order to obtain physical results and simplify the computations. In the above, one may ask whether we have not fixed the generalized lapse function $n(t,y)$ in the model, then, how it would be possible to observe physical quantities?. The answer simply lies in the fact that the dynamical EoS and deceleration parameters are not quantities that can be measured by the observers inside the brane. They belongs to bulk and are only observable by observers in the bulk. If our observations show that the universe is currently accelerating or at sometimes in future the EoS parameter crosses the phantom line, then, the model can be used to predict the position of the 3-brane in the bulk that acceleration or crossing occurs or based on the observer in the bulk at what time it might happen.

\section{numerical calculation and discussion}

From equation (\ref{eos1}), the EoS parameter varies with respect to $t$ and $y$ and for the fixed "test 3-branes" at the boundaries, the crossing and acceleration occurs at some times measured by the observer's clock in the bulk. Fig. 1 shows the 3D plot of the EoS and deceleration parameters. It shows positions in the bulk that we expect acceleration and crossing occur due to the geometry of the bulk. It also shows the at a fixed point in the bulk crossing and acceleration occurs at different times. For a better observation we take the projection of these variables onto the $y=y_0$ and $t=t_0$ planes. \\

\begin{figure}[h]
\centering
\includegraphics[scale=0.4]{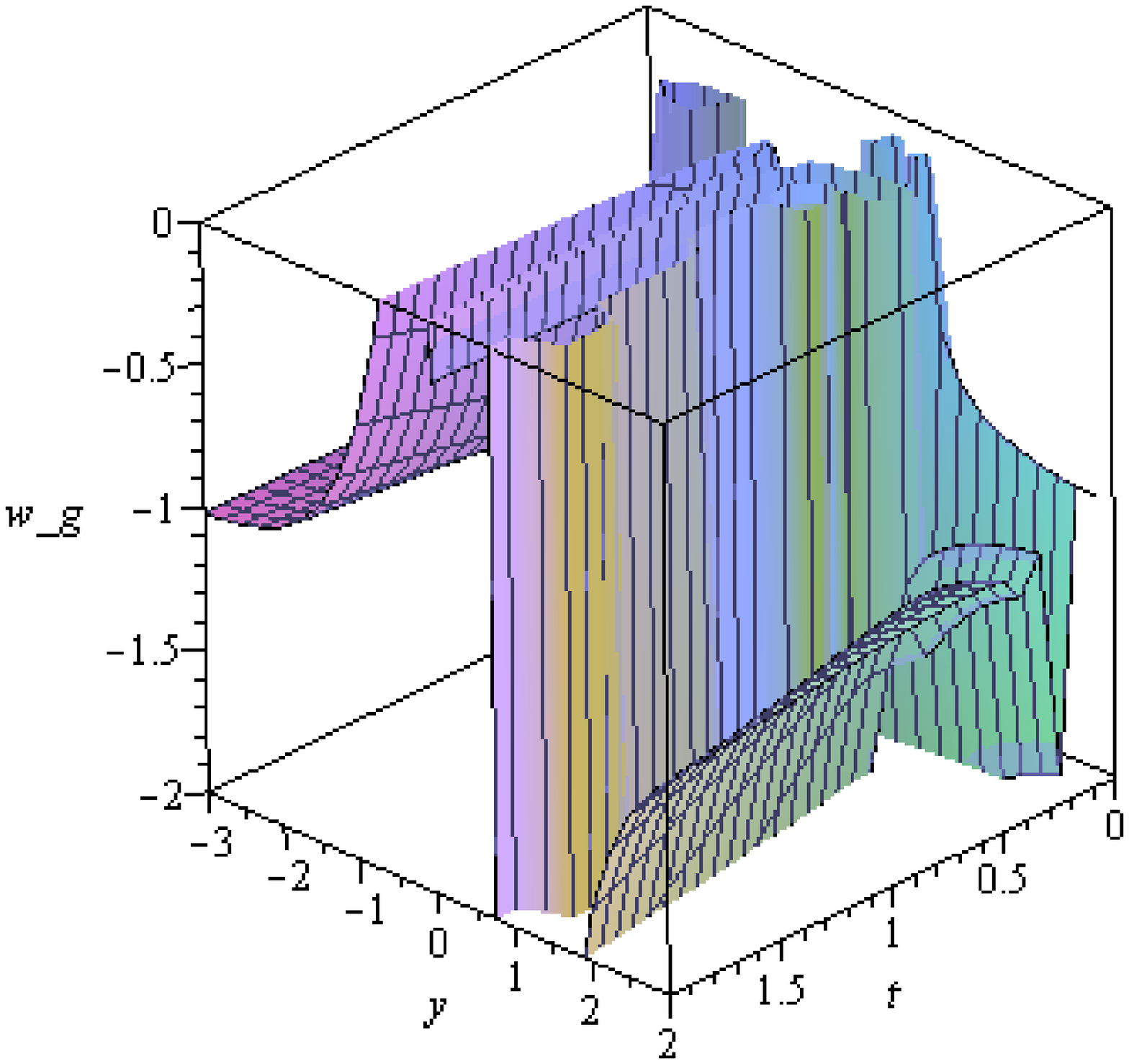}\includegraphics[scale=0.4]{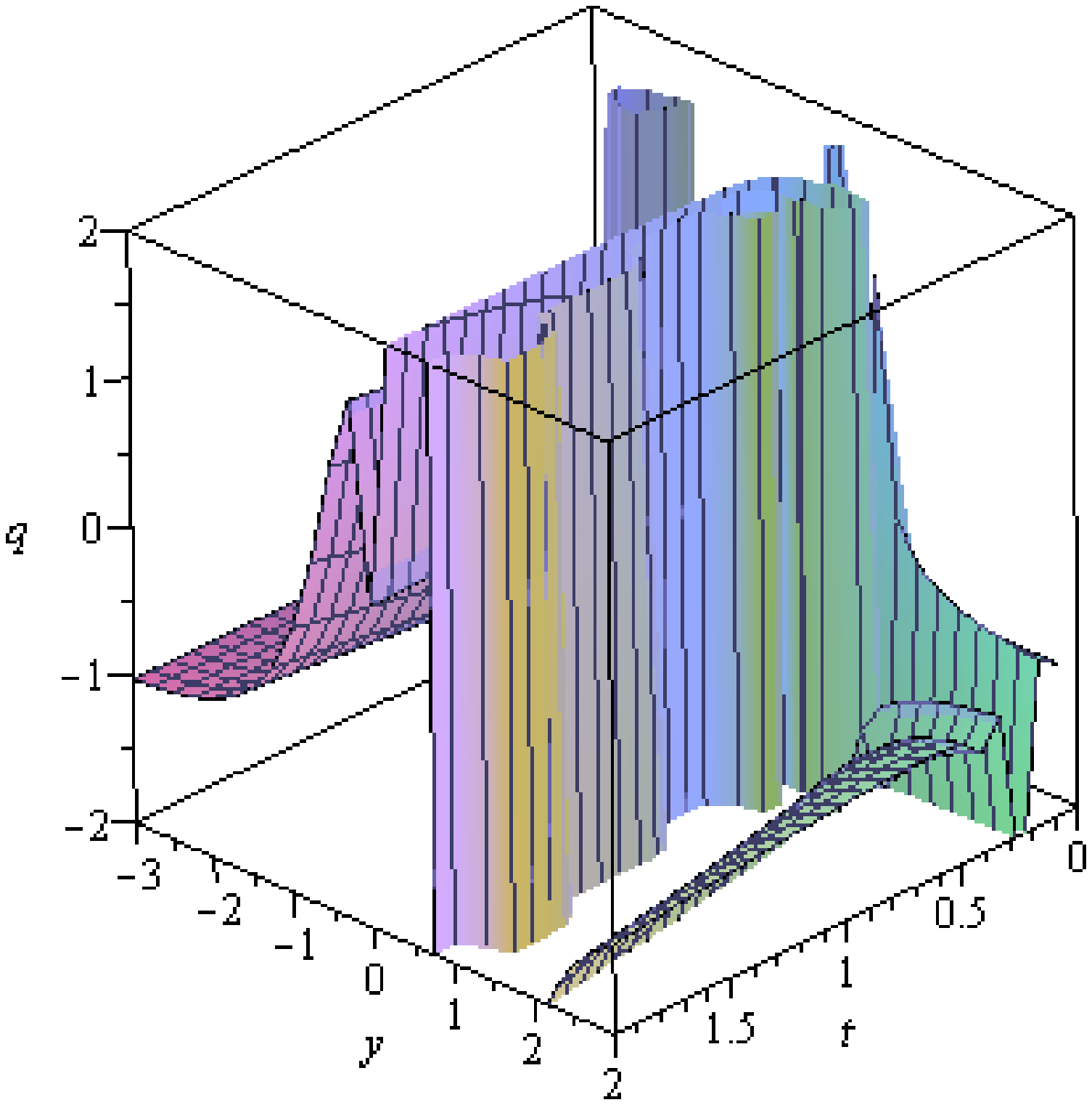}
\caption{The evolution of $\omega_{g}$ (left panel) and $q$ (right panel) with respect to $t$ and $y$.
 ICs: $a(t,0)=t^{3/5}$, $\dot{a}(t,0)=t^{1/5}$, $n(t,0)=1$.}
\end{figure}

In Fig. 2, the 2D plots of the EoS and deceleration parameters in $y$-plane are given separately for $t=t_0$. As shown, for $t=0.5$ and $t=1.5$ planes, the crossing occurs along extra dimension. Also from the graph, the deceleration parameter has similar behavior as expected. It shows that for the 3-brane located at $y=-3$, the EoS parameter is tangent to $-1$ from above at $t=0.5$. The EoS parameter is singular at a point between $y=0$ and $y=1$. In addition, in the $t=1.5$ plane, for the 3-brane located at $y=-3$, $\omega_g<-1$ and we expect crossing at $-2.5<y<-2$. The graph also shows that singularity occurs for both EoS and deceleratiopn parameters around $y=1$. From the deceleration parameter figures we see that for the 3-brane at $y=-3$, th eparameter is negative( acceleration universe), while there are points in the bulk that it is positive.

\begin{tabular*}{2.5 cm}{cc}
\includegraphics[scale=.35]{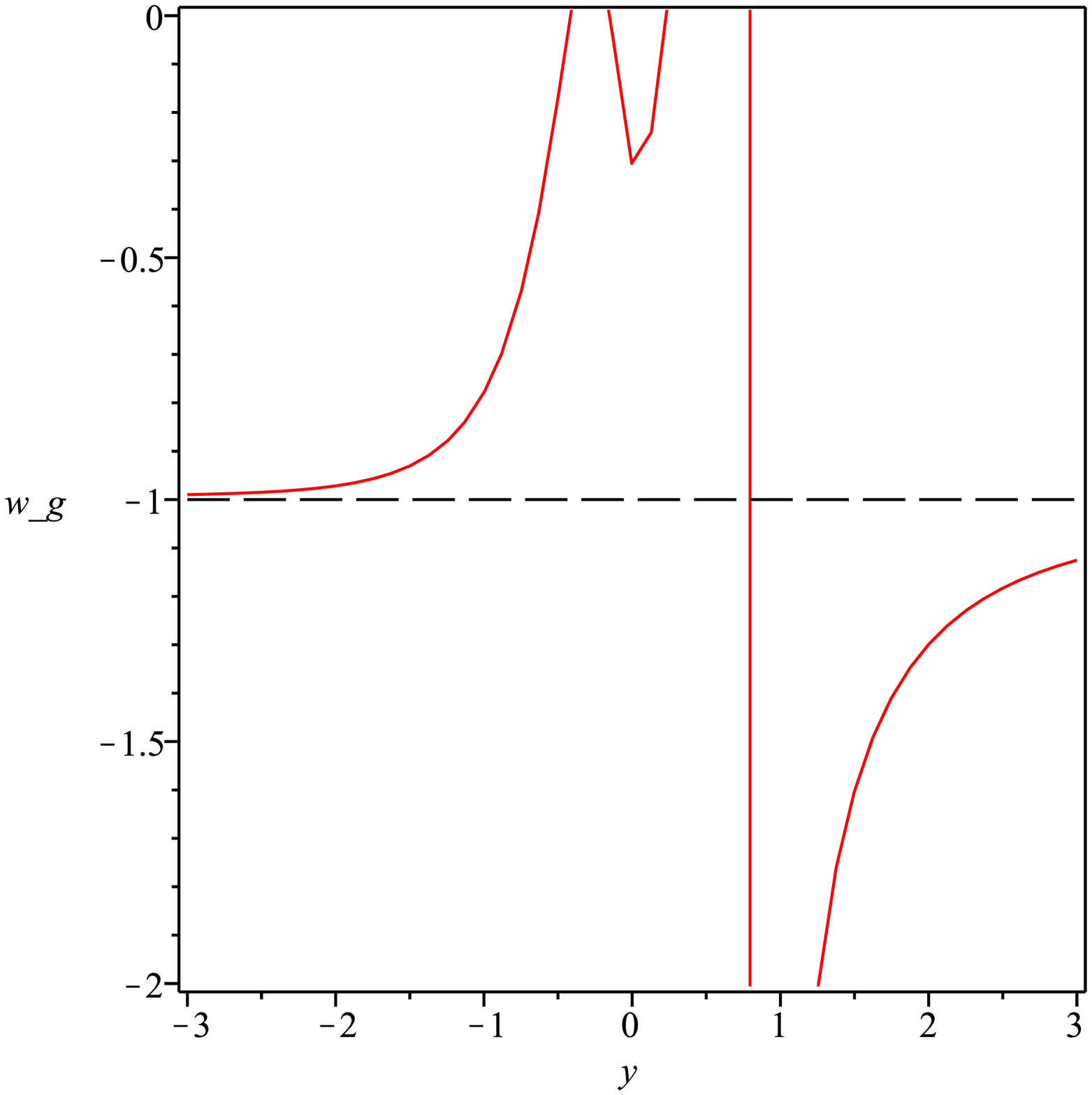} \hspace{0.1 cm}\includegraphics[scale=.35]{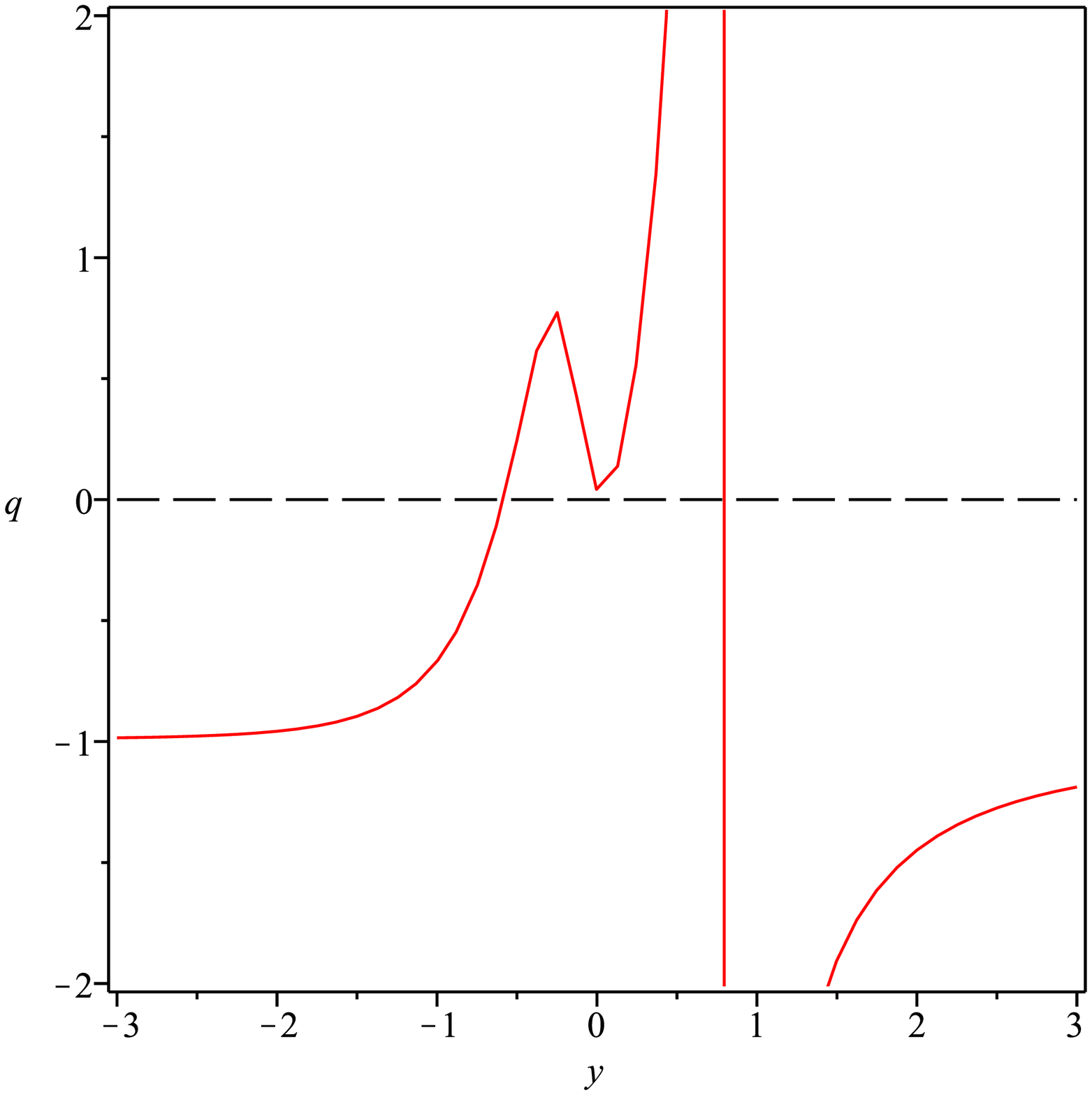}\hspace{0.1 cm}\\
\includegraphics[scale=.35]{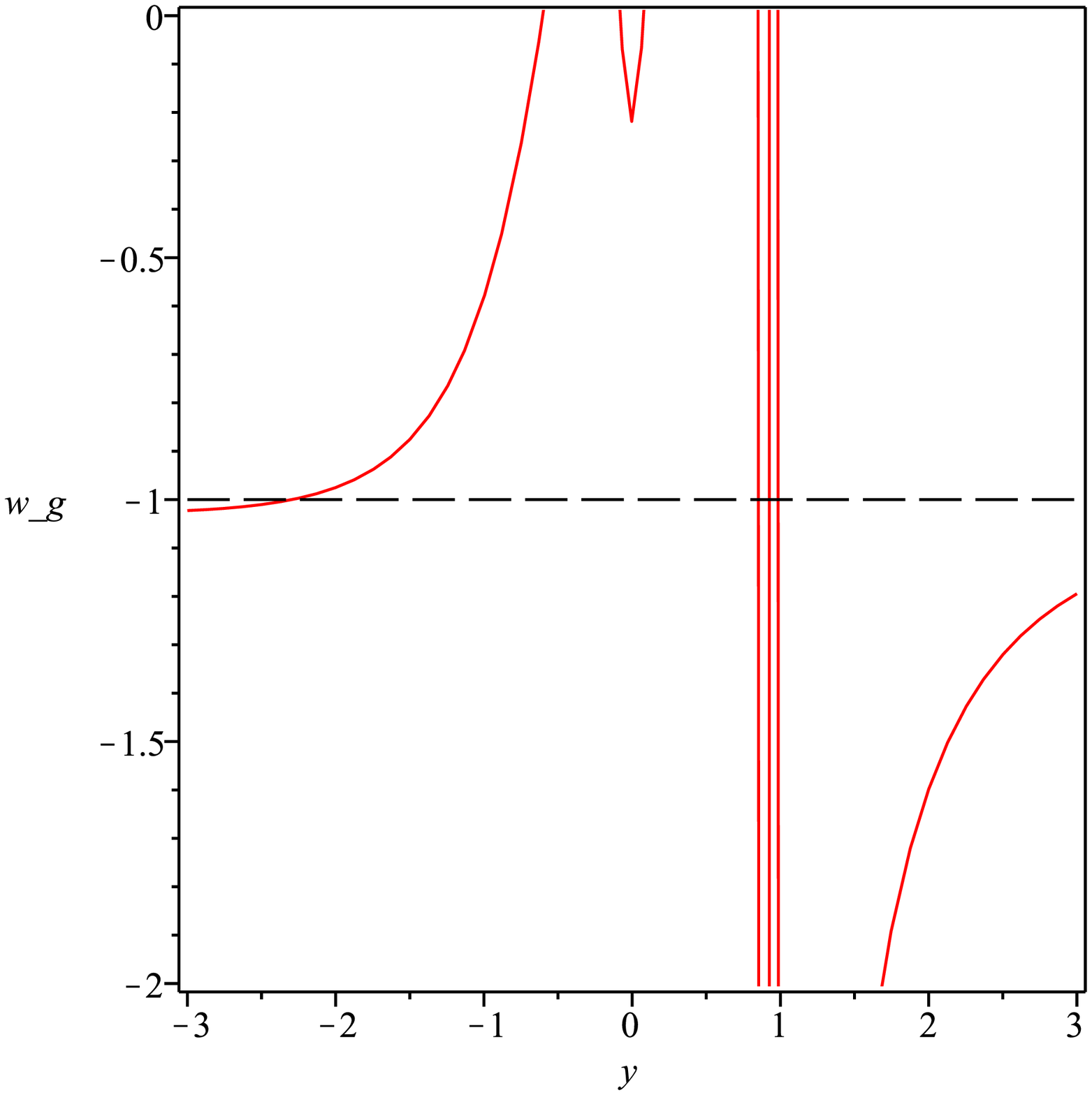} \hspace{0.1 cm}\includegraphics[scale=.35]{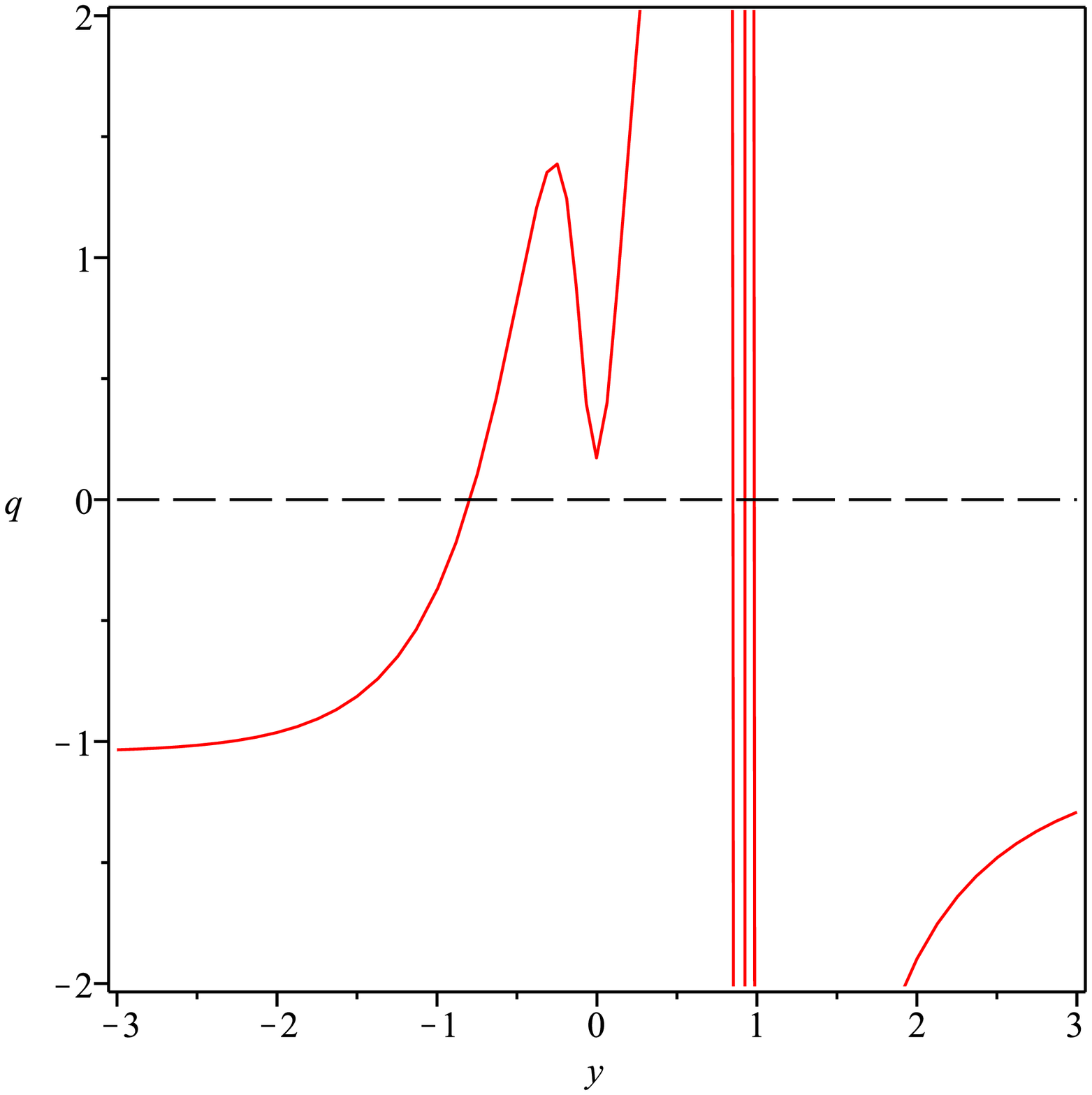}\hspace{0.1 cm}\\
FIG. 2: The projection of the the EoS parameter, $\omega_{g}$ and $q$, into $t=t_0$ planes.\\
(Top panel): $t=0.5$, (Bottom panel): $t=1.5$\\
\end{tabular*}

The projection of the EoS and deceleration parameters into the plane $y=y_0$ is shown Fig. 3. For $y=-3$, $y=0$ and $y=3$ planes, variation of EoS and deceleration parameters with time shows $\omega$-crossing and acceleration of the 3-brane. It shows that for the 3-brane located at $y=-3$, the EoS parameter is tangent to $-1$ from below at $t=0$ and from above at $t=0.5$, while crosses the line at some other times. Projection to $y=0$, shows phantom crossing at  $t=0.1$ while approaches minus infinity at $t=0$. For the 3-brane located at $y=3$, also, the crossing occurs at about  $t=0.36$, and the parameter is tangent to the line  $-1$ at  $t=0$. It is interesting to observe that the EoS parameter approaches negative infinities at some moments and at different positions in the bulk, since this is a necessary condition of realizing nonsingular bounce \cite{cai}.

\begin{tabular*}{2.5 cm}{cc}
\includegraphics[scale=.35]{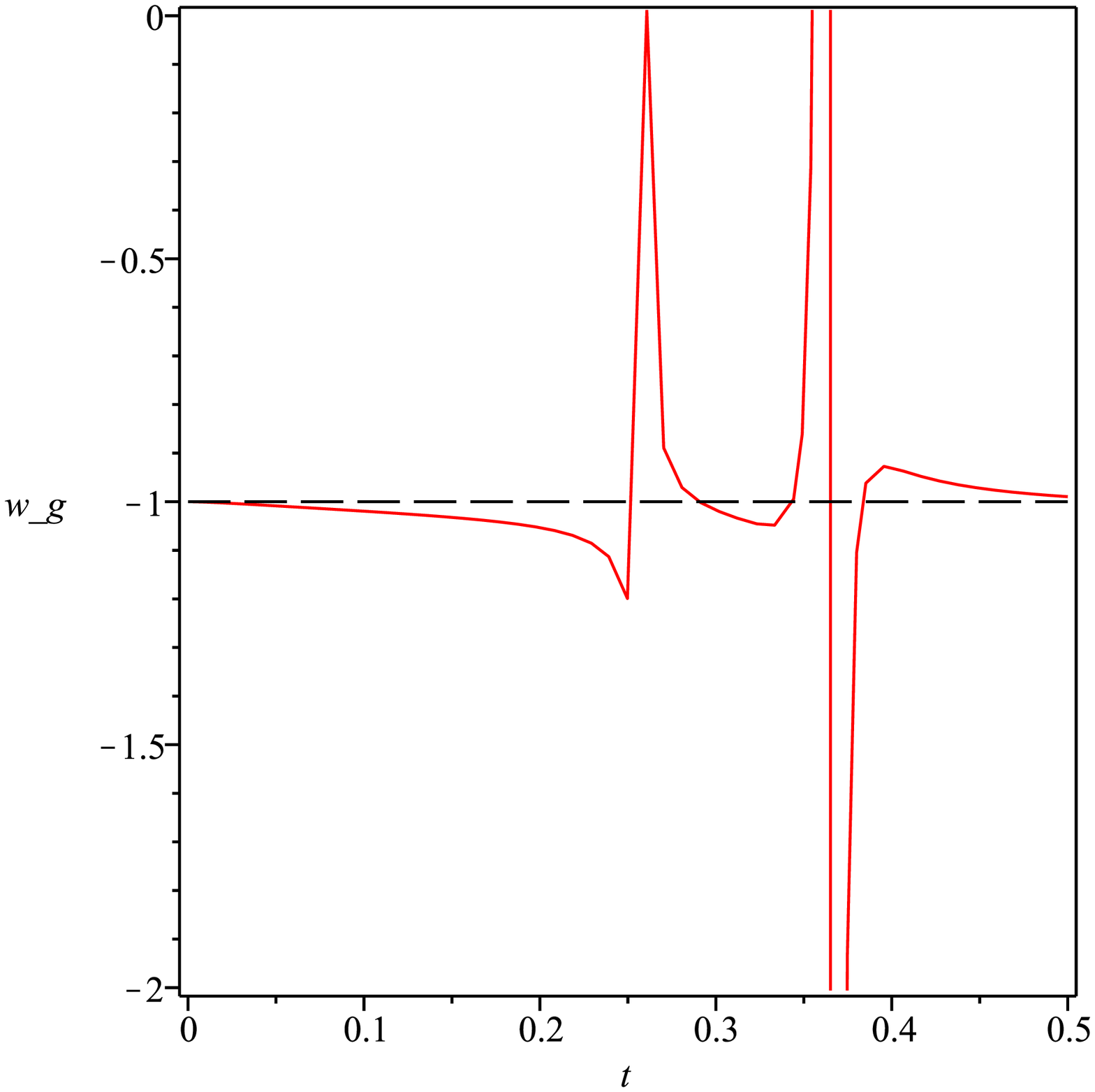}\hspace{0.1 cm}\includegraphics[scale=.35]{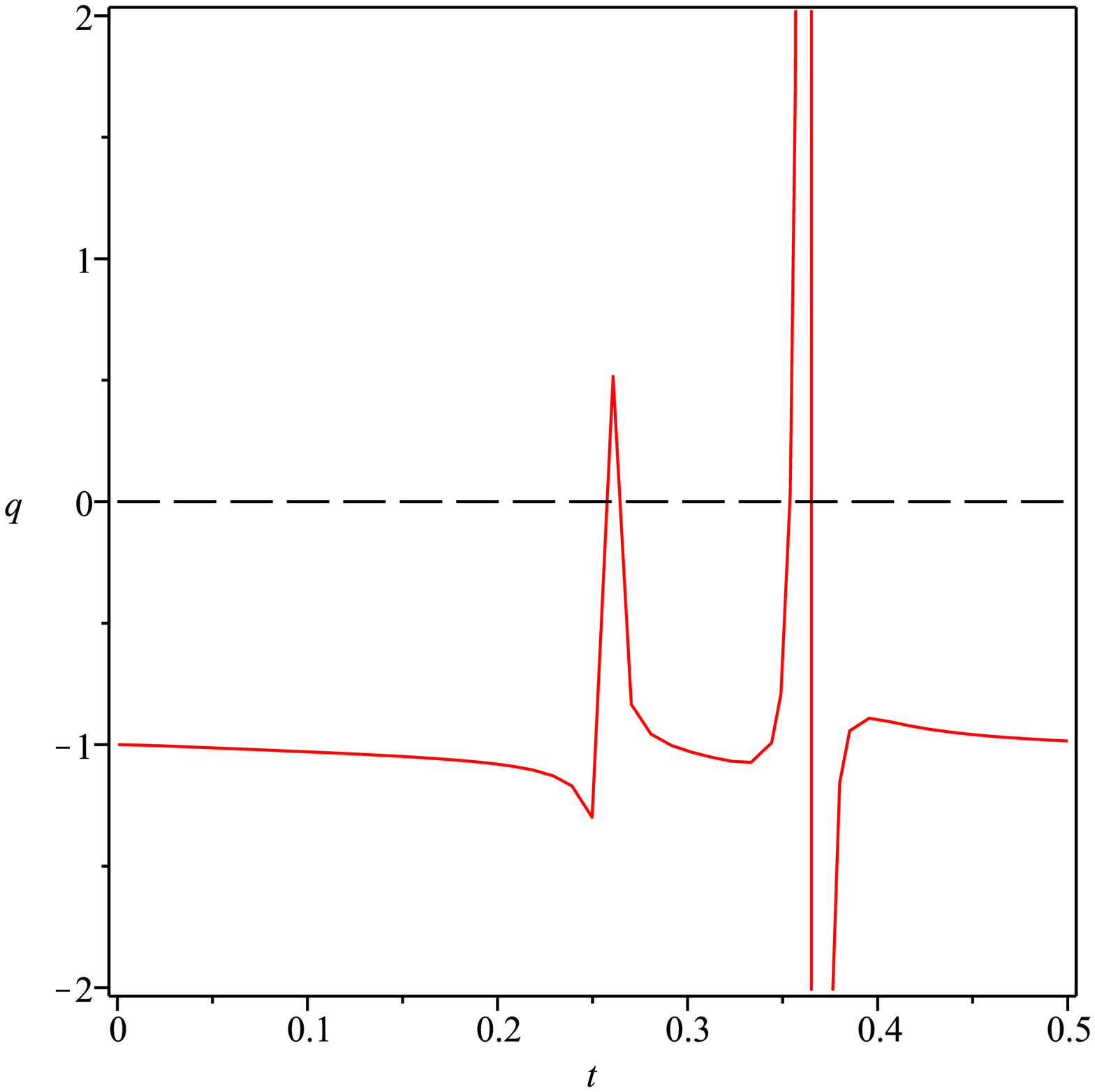}\hspace{0.1 cm}\\
\includegraphics[scale=.35]{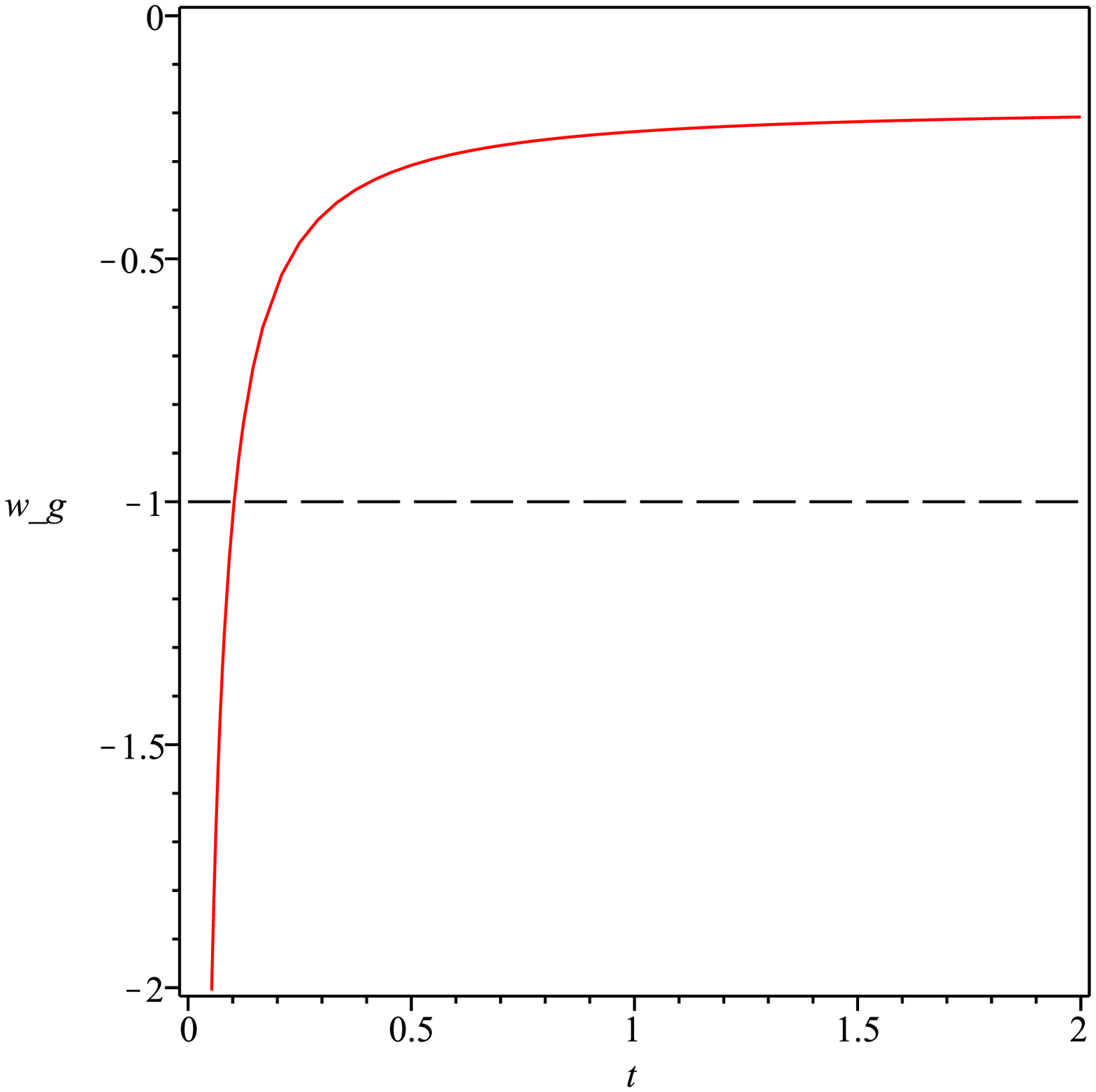} \hspace{0.1 cm}\includegraphics[scale=.35]{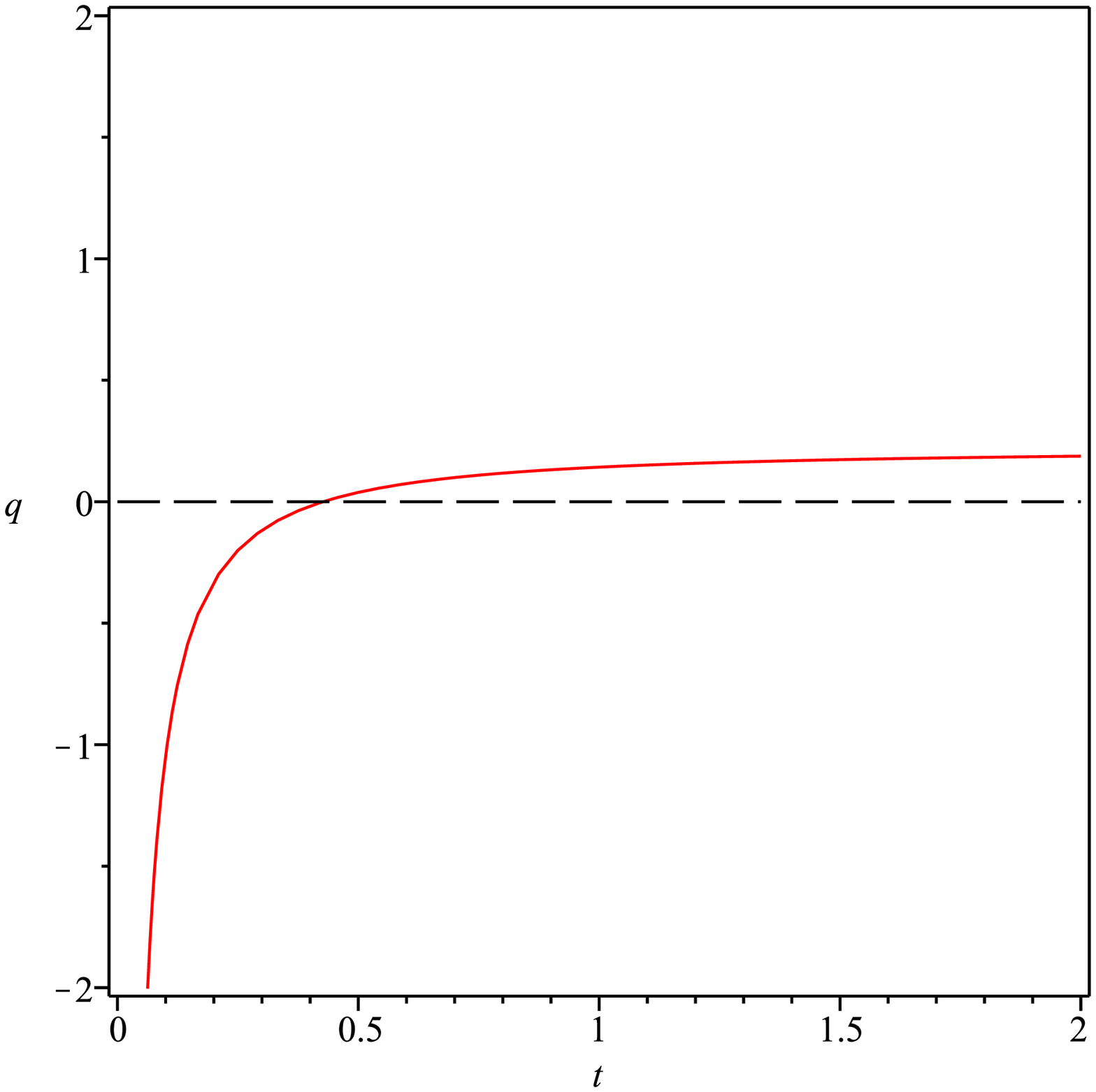}\hspace{0.1 cm}\\
\includegraphics[scale=.35]{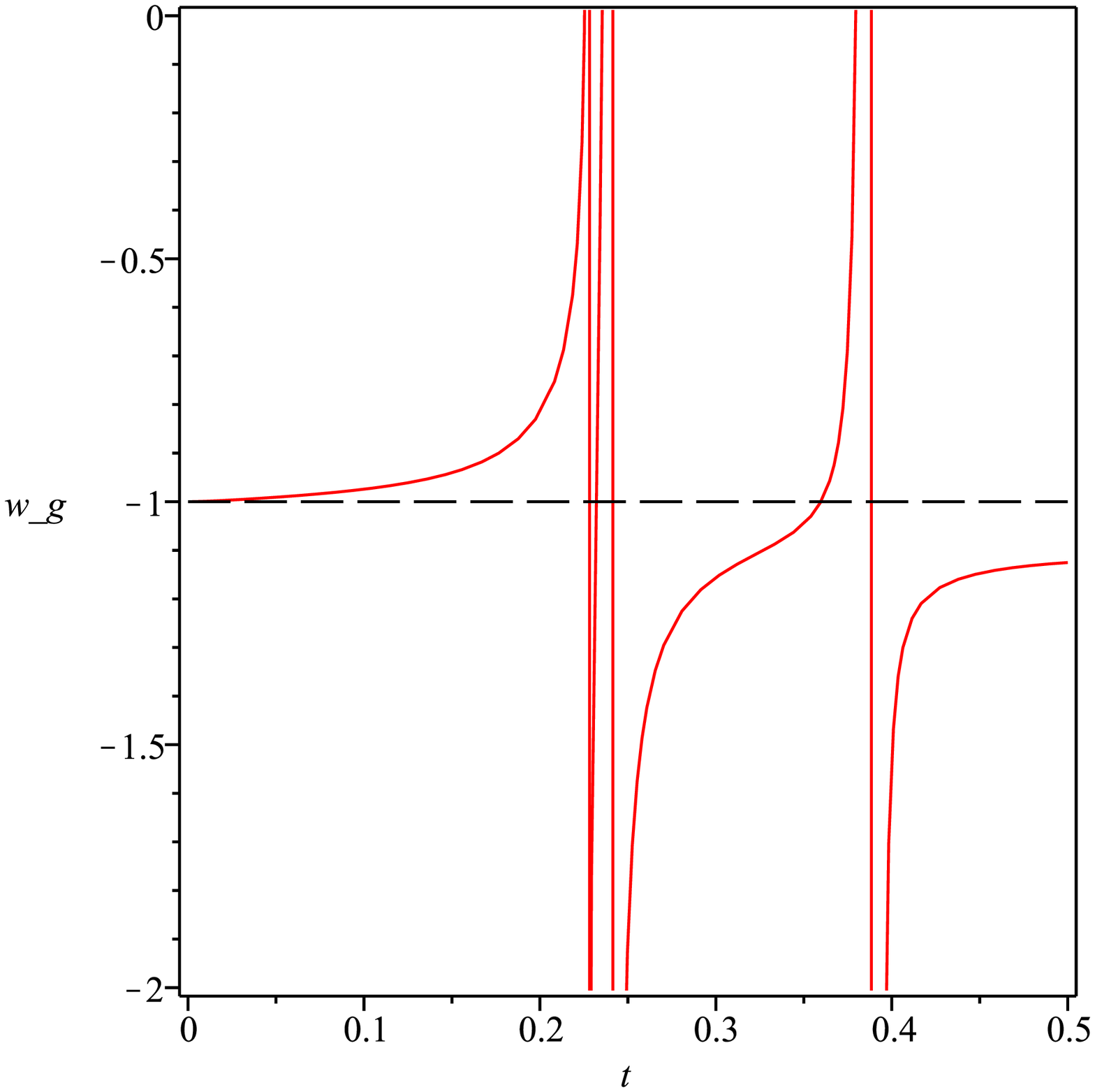} \hspace{0.1 cm}\includegraphics[scale=.35]{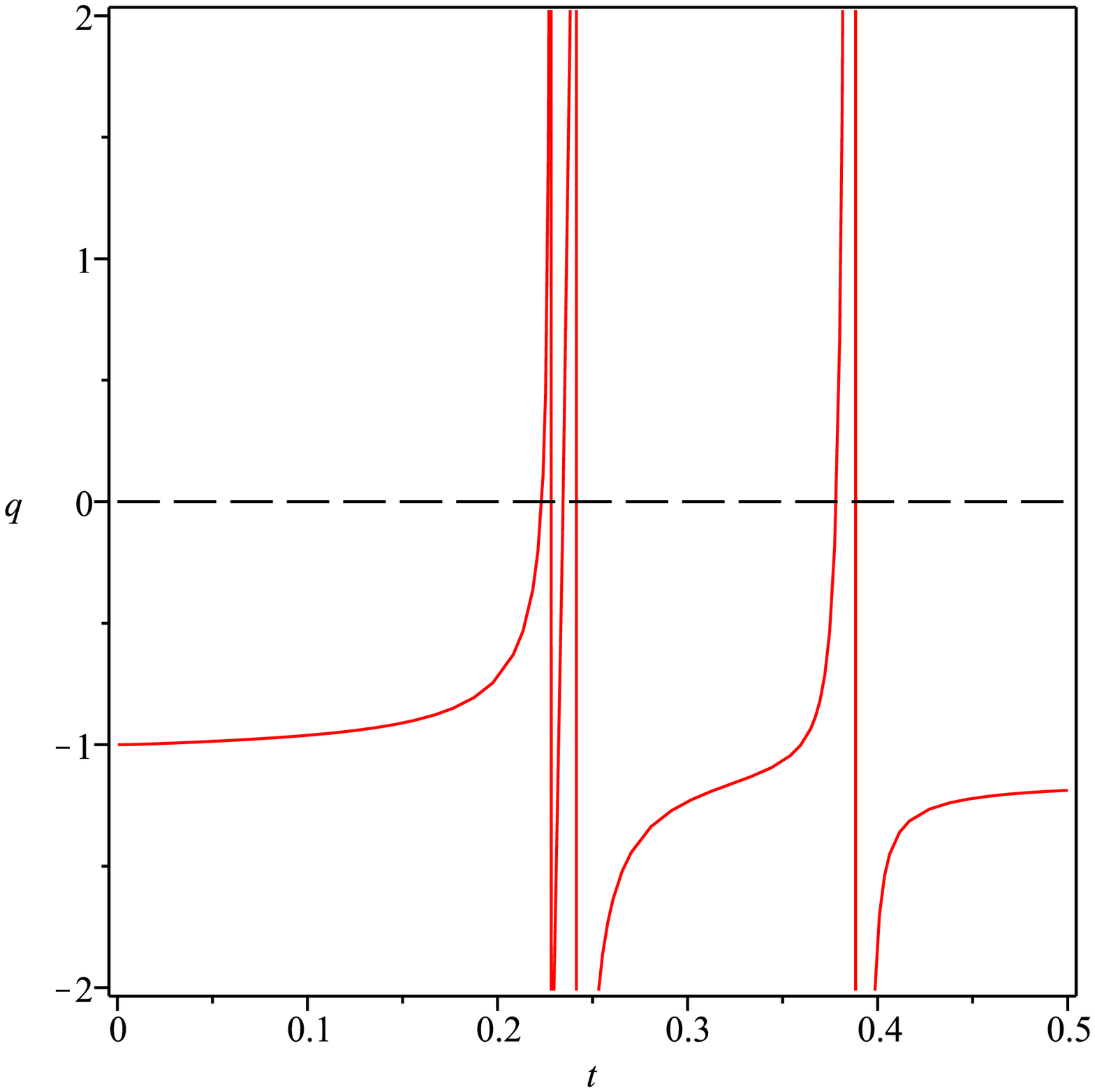}\hspace{0.1 cm}\\
FIG. 3: The projection of the the EoS parameter, $\omega_{g}$ and $q$, into $y=y_0$ planes.\\
(Top panel): $y=-3$, (Middle panel): $y=0$, (Bottom panel): $y=3$\\
\end{tabular*}

To summarize we assumed that our universe is like a "test 3-brane " inside the bulk that experience the bulk dynamical effects. This is if one neglect the 3-branes curvature, matter and tension in comparison to the bulk geometry. We drive the field equations and also the generalized EoS and deceleration parameters in the bulk. it shows that, at different points in the bulk along extra dimension and also at a fixed point but different times, phantom crossing and acceleration occur.
 In this model the "test 3-branes" are fixed in the bulk, however, one may consider the "test 3 branes" move along the bulk then again at different points in the bulk the phantom crossing and acceleration may occur.


\begin{thebibliography}{99}

\bibitem{obs00} Y. F. Cai, E. N. Saridakis, M. R. Setare and J. Q. Xia, Phys. Rep. 493, 1-60 (2010).
\bibitem{obs0} B. Feng, X. Wang and X. Zhang, Phys. Lett. B. 607, 35-41 (2005).
\bibitem{obs1} A. G. Riess et al., Astrophys. J. 607, 665-687 (2004).
\bibitem{obs11} R. A. Knop et al., Astrophys. J. 598, 102 (2003).
\bibitem{obs2} C. L. Bennett et al., Astrophys. J. Suppl. 148, 1 (2003).
\bibitem{obs3} K. Abazajian et al., Astron. J. 129, 1755 (2005).
\bibitem{obs33} M. Tegmark et al., Astrophys. J. 606, 702 (2004).
\bibitem{obs4} S. W. Allen, R. W. Schmidt, H. Ebeling, A. C. Fabian and L. van Speybroeck, Mon. Not.
Roy. Astron. Soc. 353, 457 (2004).
\bibitem{dark1} M. S. Berger and H. Shojaei, Phys. Rev. D. 74, 043530 (2006).
\bibitem{6} V. V. Kuzmichev and V. E. Kuzmichev, Ukr. J. Phys. 50, 1321 (2005).
\bibitem{dark2} J. L. Tonry et al., Astrophys. J. 594, 1 (2003).
\bibitem{dark5} S. Nojiri and S. D. Odintsov, Phys. Lett. B. 599, 137 (2004).
\bibitem{omega1} R. R. Caldwell, Phys. Lett. B. 545, 23 (2002).
\bibitem{omega11} R. R. Caldwell, M. Kamionkowski and N. N. Weinberg, Phys. Rev. Lett. 91, 071301 (2003).
\bibitem{23} I. Y. Aref'eva, A. S. Koshelev and S. Y. Vernov, Theor. Math. Phys. 148, 895 (2006).
\bibitem{omega2} L. A. Boyle, R. R. Caldwell and M. Kamionkowski, Phys. Lett. B.545, 17-22 (2002).
\bibitem{cross1} L. P. Chimento, R. Lazkoz, R. Maartens and I. Quiros, JCAP. 0609, 004 (2006).
\bibitem{omega4} D. J. Eisenstein et al., Astrophys. J. 633, 560-574 (2005).
\bibitem{eos2} U. Seljak et al., Phys. Rev. D. 71, 103515 (2005).
\bibitem{eos22} M. Tegmark, JCAP. 0504, 001 (2005).
\bibitem{eos3} M. R. Setare and J. Sadeghi, Int. J. Theor. Phys. 47, 3219 (2008).
\bibitem{eos4} E. N. Saridakis and J. M. Weller, Phys. Rev. D. 81, 123523 (2010).
\bibitem{mtheory7} M. Bouhmadi-Lopez, Nucl. Phys. B. 797, 78-92 (2008).
\bibitem{mtheory} N. Agarwal, R. Bean, J. Khoury and M. Trodden, Phys. Rev. D. 81, 084020 (2010).
\bibitem{mtheory3} H. M. Lee and G. Tasinato, JCAP. 0404, 009 (2004).
\bibitem{mtheory4} D. Karasik and A. Davidson, Class. Quant. Grav. 21, 1295-1302 (2004).
\bibitem{mtheory5} K. Aoyanagi and K. Maeda, JCAP. 0603, 012 (2006).
\bibitem{mtheory6} Y. Shtanov, A. Viznyuk and V. Sahni, Class. Quant. Grav. 24, 6159-6190 (2007).
\bibitem{acc1} I. Quiros, R. Garcia-Salcedo and C. Moreno, Phys. Rev. D. 75, 023510 (2007).
\bibitem{acc2} J. S. Alcaniz and N. Pires, Phys. Rev. D. 70, 047303 (2004).
\bibitem{mod1} M. B. Lopez and A. Ferrera, JCAP. 0810, 011 (2008).
\bibitem{brane1} G. Kofinas, G. Panotopoulos and T. N. Tomaras, JHEP. 0601, 107 (2006).
\bibitem{brane4} S. Lepe, F. Pe$\check{n}$a and J. Saavedra, JHEP. 0805, 042 (2008).
\bibitem{brane2} P. Binetruy, C. Deffayet and D. Langlois, Nucl. Phys. B. 565, 269 (2000).
\bibitem{17} P. Binetruy, C. Deffayet, U. Ellwanger and D. Langlois, Phys. Lett. B. 477, 285 (2000).
\bibitem{brane3} J. M. Cline, C. Grojean and G. Servant, Phys. Rev. Lett. 83, 4245 (1999).
\bibitem{rs} L. Randall and R. Sundrum, Phys. Rev. Lett. 83, 3370 (1999).
\bibitem{bra1} K. Nozari, M. R. Setare, T. Azizi and N. Behrouz, Phys. Scripta. 80, 025901 (2009).
\bibitem{bra2} P. S. Apostolopoulos, N. Brouzakis, E. N. Saridakis and N. Tetradis, Phys. Rev. D. 72, 044013 (2005).
\bibitem{crossing} E. N. Saridakis, Phys. Lett. B. 661, 335 (2008).
\bibitem{crossing1} E. N. Saridakis, Nucl. Phys. B. 830, 374 (2010).
\bibitem{crossing2} P. Moyassari and M. R. Setare, Phys. Lett. B. 674, 237 (2009).
\bibitem{bounce1} S. Foffa, Phys. Rev. D. 68, 043511 (2003).
\bibitem{bounce2} T. J. Battefeld, S. P. Patil and R. Brandenberger, Phys. Rev. D. 70, 066006 (2004).
\bibitem{bounce3} H. Zhang and Z. H. Zhu,  Phys. Rev. D. 75, 023510 (2007).
\bibitem{israel2} A. Balcerzak and M. P. Dabrowski, Phys. Rev. D. 81, 123527 (2010).
\bibitem{crs} K. Enqvist, E. Keski-Vakkuri and S. Rasanen., Phys. Rev. D. 64, 044017 (2001).
\bibitem{Kanti} P. Kanti, K. A. Olive and M. Pospelov, Phys. Rev. D. 62, 126004 (2000).
\bibitem{Brevik} I. Brevik, K. B{\o}rkje and J. P. Morten, Gen. Rel. Grav. 36, 2021-2038 (2004).
\bibitem{Cline} J. M. Cline and J\'{e}r\'{e}mie Vinet, AIP Conf. Proc. 646, 209-216 (2003).
\bibitem{Arnowitt} R. Arnowitt, J. Dent and B. Dutta, Phys. Rev. D. 70, 126001 (2004).
\bibitem{Kim} N. J. Kim, H. W. Lee, Y. S. Myung and G. Kang, Phys. Rev. D. 64, 064022 (2001).
\bibitem{Mersini} L. Mersini, Mod. Phys. Lett. A. 16, 1583-1596 (2001).
\bibitem{Leon} J. P. de Leon, Gen. Rel. Grav. 36, 923-948 (2004).
\bibitem{csaki} C. Csaki, M. Graesser, L. Randall and J. Terning, Phys. Rev. D. 62, 045015 (2000).
\bibitem{Lesgourgues} J. Lesgourgues, S. Pastor, M. Peloso and L. Sorbo, Phys. Lett. B. 489, 411 (2000).
\bibitem{Mohapatra} R. N. Mohapatra, A. P\'{e}rez-Lorenzana, C. A. de S. Pires, Int. J. Mod. Phys. A. 16, 1431-1442 (2001).
\bibitem{cai} Y. F. Cai, T. Qiu, R. Brandenberger and X. Zhang, Phys. Rev. D. 80, 023511 (2009).

\end{thebibliography}
\end{document}